\documentstyle[12pt]{article}

\begin{document}
\noindent
\begin{center}
\begin{large}
{\bf Local angular fractal and galaxy distribution
}\\
 \end{large}
Yu Shi\footnote{Email: shiyu@alon.cc.biu.ac.il}\\
Department of Physics, Bar-Ilan University, Ramat-Gan 52900, Israel
\\ 
\end{center}

{\bf Abstract}

The power-law dependence of the angle in the angular projection of 
galaxy distribution is explained by assuming that in the spherical shells
 within
a small angle the distributions are also fractal. If  this local 
angular fractal 
is possessed, a fractal structure is
angularly-isotropic at each occupied point though inhomogeneous, and
is compatible with the present evidence claimed to be of homogeneity for
galaxy distribution. Further, it is most likely to be isotropic rather than
only angularly-isotropic. 
Several related issues are discussed.    

\vspace{1.5cm}

$PACS$ : 98.62.Py, 98.80.Es, 05.90.+m, 61.43.Hv

$Keywords$: Fractal; Galaxy distribution; Universe
 
\newpage
 The past two decades witnessed an explosive activity on the scale-invariant 
 but non-homogeneous objects, fractals \cite{Mandelbrot}, in almost all 
 branches of science. Along with the findings of more structures by redshift
 surveys, it is interesting and important to know whether the large scale 
 structure of the universe is a fractal, a question which was actually a 
 stimulation in developing fractal geometry \cite{Mandelbrot}.
 While it has been a consensus that the galaxy distribution approximates
  a fractal  over a considerable scales\cite{Peebles},
   it is  still under 
   debate whether it crossovers to homogeneity on a scale about, say, 
   $20h^{-1}Mpc$ \cite{Borgani}
  \cite{Davis}, or extends up to the present
  limit of observation as claimed by Pietronero and 
  his collaborators (hereafter P)\cite{Pietronero}\cite{CP}.
A crucial  problem in this context concerns
the angular projection of the galaxy distribution
since before the extensive redshift survey the
    galaxy catalogs were for 
    angular coordinates and now there has been 
 much angular information  \cite{Davis}. 
    Based on a numerical simulation on a fractal structure geneated by a
 Levy flight in 3-dimensional space, 
 P stated that the angular projection shows fractal correlation at small 
 angles  but becomes homogeneous at large angles
 \cite{CP}, a so-called local isotropy is claimed to exist and realize
  the Cosmological Principle in the fractal structure \cite{Pietronero}. 
 As we point out later, however, their approach contains an
  implicit assumption of angular 
 homogeneity in {\em distribution}.
 Note that angular distribution and projection are two different 
 concepts. 
  It will be shown in a consistent way that power laws for angular 
  projection at small angles are manifestations of 
   fractal of
  angular  {\em distribution}.
  The bahavior at
  large angles deserves further study.
  We distinguish homogeneity from isotropy, which is also different from
  only angular-isotropy. 
 A fractal structure with local
  angular fractal is  inhomogeneous but angularly-isotropic, hence can
   satisfy much evidence calimed to be of homogeneity
    but actually of angular-isotropy.
   Generally a fractal can be  neither homogeneous nor angularly-isotropic, 
   cosequently, as we will see, the evidence of fractal on scales larger
    than 
 $50h^{-1}Mpc$ is quite problemic. However, they will be more evident 
 on the contrary if  some basis of opposition arguments \cite{Davis}
 is established.  
   
  In the traditional  analysis, one usually calculates the two-point correlation
 function
 \begin{eqnarray}
 \xi(r) &=& \frac{<n(\stackrel{\rightarrow}{r}_{0})\cdot
 n(\stackrel{\rightarrow}{r}_{0}+\stackrel{\rightarrow}{r})>_{\stackrel
 {\rightarrow}{r}_{0}}}{<n>^{2}}-1 
 \nonumber
 \\
  &=& \frac{<n(\stackrel{\rightarrow}{r}_{0}+\stackrel{\rightarrow}{r})>
 _{\stackrel{\rightarrow}{r}_{0}}}{<n>}-1,	\label{eq:xi}
 \end{eqnarray}
and the the angular correlation function
 \begin{eqnarray}
 \omega(\theta)&=& \frac{<n(\theta_{0})\cdot n(\theta_{0}+\theta)>
 _{\theta_{0}}}{<n>^{2}}-1	   \nonumber\\
 &=& \frac{<n(\theta_{0}+\theta)>_{\theta_{0}}}{<n>}-1, \label{eq:omega}
 \end{eqnarray}
 where $n(\stackrel{\rightarrow}{r}_{0}+\stackrel{\rightarrow}{r})$
 is the number density, 
 $<\,>_{\stackrel{\rightarrow}{r}_{0}}$ means average over 
settings of the origin 
$\stackrel{\rightarrow}{r}_{0}$ over the sample.
$<n>$ 
 is the average density within the sample. The angular correlation function 
 is defined in the similar way for angular coordinates, but note that 
 densities are defined 
 for the solid angle $2\pi(1-\cos \theta)$ instead of $\theta$
 since the sample 
 is a conic part of a sphere. This makes the angular 
 projection more complicated than the original distribution.
 Since within the crossover scales
 for homogeneity, the average density and therefore the correlation functions
 are dependent on the sample depth, P suggested to study particularly
 the conditional densities $\Gamma(r)$ and $\Gamma(\theta)$,
 which are just the numerators of the first terms in rhs.
 of Eqs. (\ref{eq:xi}) and (\ref{eq:omega}), respectively. 
 
 The Limber
  scheme gives \cite{Peebles}\cite{Borgani}
   $\omega(\theta)\,\propto\,(r_{0}/L)^{\gamma}\theta^{1-\gamma}$, where
   $L$ is the sample depth, the length of the cone, 
   $r_{0}$ is defined by $\xi(r_{0})\,=\,1$ 
   therefore is proportional to $L$ \cite{CP}.
   As pointed out by P in their pionnering work, the homogeneity
   assumption was taken in Limber scheme, as well as in the manipulation of
    data, without considering the possible dependence on the sample 
    angle $2\theta_{M}$. Therefore the consistency of analysis with
      homogeneity
  claiming  that $\omega(\theta)$ for galaxy distribution
  is dependent on sample depth \cite{Borgani}  is questionable, also
 we cannot deduce the functions for the case of 
 fractal just by setting $r_{0}\,\propto\,
 L$ in that obtained through Limber scheme.
  According to P, for
   both the angular data from galaxy catalogs and the
 fractal generated by a Levy flight, $\omega(\theta)$ increases with
 $\theta_{M}$ while independent of L.  
    From Fig. 36 in Ref. \cite{CP} 
  we may observe that the feature of $\omega(\theta)$ 
  is almost the same as $\xi(r)$ (Figs. 5 and 22 there).
    As exposed later, this is due to the local angular fractality. 
    $\Gamma(\theta)$ for CfA was found to increase with the depth.
   Both $\omega(\theta)+1$ for simulated fractal  and
   $\Gamma(\theta)$ for CfA 
   depend on
   $\theta$ with a power law when $\theta$ is small.
      
For a fractal distribution, the number of points within a sphere of a radius
$r$ is $N(r)\,=\,Br^{D}$, where D is the fractal dimension, B is a 
coefficient. A crucial concept thrughout the work of P is the
 conditional density from
an occupied point defined as \cite{CP}
\begin{equation}
\Gamma(r)\,=\,\frac{1}{S(r)}\frac{dN(r)}{dr}\,=\,(\frac{DB}{4\pi})r^{D-3},
\end{equation}
where $S(r)$ is the area of a spherical shell of radius $r$. 
We note that this is a quantity averaged over all the angular directions,
hence loses
information on angular distribution.  This issue is  irrelevent 
when one only determine $D$  sampling with sphere. But there are two 
approaches where $\Gamma(r)$ was inappropriately used, taking the 
implicit assumption
that it is homogeneous on the sphere, {\it i.e.}, angular 
homogeneity. One is that to the angular correlation,
 explaining the dependence of $\theta^{*}$
on $\theta_{M}$ \cite{CP}, here $\theta^{*}$ is defined by 
$\omega(\theta^{*})\,=\,1$. The number of points on a spherical shell 
within $\theta$ was taken as $\Gamma(r)\Omega(\theta)r^{2}$. So it is 
inconsistent in a way
similar to what they criticized.
In 
fact, the dependence of $\Gamma(\theta)$ and $\omega(\theta)$ on $\theta$ 
is just
 a signature of angular inhomogeneity in distribution.
 In the following the behavior of  angular projection is explained
  in a simple and natural way by 
 the so-called local angular fractal. At first it is an assumption
and approximation, nevertheless it is much better than assuming 
angular homogeneity, its validity is proved by the results and is 
expected to 
be established in a more rigorous formulation under way, on the other hand it
may be tested on the simulated fractal and the data of galaxy distribution. 

 Consider a small conic part of a sphere defined by  $L$ and $2\theta$.
 The local angular fractal referrs to that 
  when $\theta$ is small enough, the points on each spherical shell within 
  this conic part have fractal distribution while the spherical shells 
  also distribut fractaly. So actually, it is a sort of bifractal \cite{Shi},
  a product of two subfractals, locally. In many cases such as 
  galaxy distribution, it should be, of course, in a statistical way. 
  The number of points in the considered volume is 
  \begin{equation}
 N(L,\theta)\,=\,AD\int (r\theta)^{D_{\Omega}}r^{D_{r}-1} dr\,=\
 AL^{D}\theta^{D_{\Omega}},	     \label{eq:rt}
 \end{equation}
 where $A$ is a coefficient, $D_{r}$ and $D_{\Omega}$ are the radial and 
 angular dimensions respectively, with $D\,=\,D_{r}\,+\,D_{\Omega}$. 
 If the sample size is characterized by the angle $2\theta_{M}$ and the 
 radial  depth $L$, then the solid angle is $\Omega(\theta_{M})\,=\,
 2\pi(1-\cos \theta_{M})\,\approx\,\pi\theta_{M}^{2}$,
the average angular density is thus
 $<n>_{M}\,=\,N(\theta_{M})/\Omega(\theta_{M})\,
 \approx\,(A/\pi)L^{D}\theta_{M}^{D_{\Omega}-2}$.
 The density defined at the origin 
 is $n(\theta)\,=\,dN(\theta)/d\Omega(\theta)\,=\,$
 $(dN/d\theta)/(d\Omega/d\theta)\,\approx\,
 (AD_{\Omega}/2\pi)L^{D}\theta^{D_{\Omega}-2}$, 
 theoretically it is just $\Gamma(\theta)$.
 The angular correlation function is 
 $\omega(\theta)\,=\,\Gamma(\theta)/<n>_{\theta_{M}}-1$$\,
 \approx\,(D_{\Omega}/2)(\theta/\theta_{M})^{D_{\Omega}-2}-1$. 
 Similar to the conditional average
  density \cite{CP}, we may also obtain a conditional average angular
  density  $\Gamma^{*}(\theta)\,=\,N(\theta)/\Omega(\theta)\,\approx\,
  (A/\pi)L^{D}\theta^{D_{\Omega}-2}$. For $D_{\Omega}\,<\,2$, it is 
  obvious that 
   $\omega(\theta)$ increases with $\theta_{M}$ and dependent of $L$,
  while $\Gamma(\theta)$ is independent of $\theta_{M}$ and increases with 
  $L$.
$\theta^{*}$  defined by  $\omega(\theta^{*})\,=\,1$ is given by 
$\theta^{*}\,\approx\,(D_{\Omega}/4)^{1/(2-D_{\Omega})}\theta_{M}$, it is
as spurious as $r_{0}$, since we may see that $\omega(\theta)$ is not a
well defined function for angular inhomogeneity. 
To see this, compare 
(a) for $\omega(\theta)$ and (b) for $\Gamma(\theta)$ in Fig. 1.
Since $\Gamma(\theta)$ 
and $\Gamma^{*}(\theta)$ are independent of $\theta_{M}$, they can be used
to study angular distribution at small angles. For large angles, unlike
the corresponce in the full distribution,
the solid angle is not simply a power function of $\theta$, things become
artificially complicated as observed. The best way
is just to study the scaling of the 
number of points (galaxies) with the angle. 
   
  So the power laws are direct consequences of local angular fractal.
  There was, of course, no direct investigation on $D_{\Omega}$. It can be 
  estimated to be $1.3$ for $\omega({\theta})\,\propto\,\theta^{-0.7}$
  \cite{CP}. This is
  consistent with the finding that $D\,\approx\,2$ \cite{Pietronero}, which
  implies that $D_{\Omega}$ should be larger than $1$ and less than $2$.
  It should be pointed out that Eq. (\ref{eq:rt}) cannot be generalized to
  an arbitrary angle. This can be understood by considering the case of 
  homogeneity whch takes $D_{\Omega}\,=\,2$, the volume is proportional 
  to $(1-\cos\theta)$ rather than $\theta^{2}$.  The property at
  large angles deserves further mathematical investigation, which is 
  undertaken.

  At this stage, we stress the difference between angular distribution and 
  angular projection, as well as that among isotropy, angular-isotropy
   and angular homogeneity.
  The angular distribution referrs to the number of points as  a function of 
  angle within a given radial depth.
  If it is homogeneous, then $N(\theta)\,\propto\,\theta^{2}$ at small angles.
   If it is 
  angular fractal, then $N(\theta)\,\propto\,\theta^{D_{\Omega}}$ 
  ($D_{\Omega}\,<\,2$) at
  small angles. For a fractal
  in the whole space, there exist four possibilities. (i) Angular and radial 
  irregular inhomogeneity. By irregular, we mean  there is no fractal
  scaling.  
    (ii)Radial fractal and local angular fractal.
  (iii) Radial homogeneity and local angular fractal, this is a special case
  of (ii) with $D_{r}\,=\,1$ and $D\,=\,D_{\Omega}+1$. 
  (iv) Radial fractal and angular
  homogeneity. In this case, one may see that not all points are statistically
  equivalents, there is one or more than one centers, the total dimension is 
  larger than 2. For homogeneity in the whole space, it is, of course,
  homogeneous both radially and local-angularly. Angular-isotropy 
  referrs to
 that there are same number of points within same degree
  of angle, or say the angualr distribution is isotropy. 
  So not only angular homogeneity, but also local angular fractal
is angularly-isotropic. Only if radial direction and angular direction 
is also equivalent, i.e., $D_{r}\,=\,D_{\Omega}/2$, the structure is not
 only angularly-isotropic but also isotropic.
Now let us turn to the angular projection.
 It is defined through {\em solid-angular} densities {\em represented as functions 
 of $\theta$}, so is more or less artificial. A homogeneous distribution in $3d$
 space will, of course, lead to a homogeneous angular projection. But 
 {\em in principle}, as being realized by P, there may be artificial
 homogeneity, which mean the value of the function does not change with 
 $\theta$. But we are still uncertain about  the angular projection of
 a fractal for an arbitrary  angle.
However, it is sure that 
for a fractal with local angular fractal, no matter whether 
the angular projection is homogeneous,
the angular {\em distribution}, which is directly physical, 
is angularly-isotropic though not homogeneous. To investigate angular distribution, 
as pointed out above, one had better just  study the 
number of points (galaxies) within the corresponding  angle. 

It was claimed that local isotropy can exist in a fractal and realize the 
Cosmological Principle \cite{Pietronero}.
 We see now this fractal should possess local 
angular fractal. If the universe is a fractal but not with local 
angular fractal, 
 there cannot be local isotropy at every point, and the Cosmological 
Priciple relaxes just to the statistical equivalence of each point.
 Similar 
situation is in the evidence of fractal above $\sim150h^{-1}Mpc$
\cite{Pietronero}, which
is another approach by appropriately using $\Gamma(r)$.
To use  thin deep catalogues up to the total depth, which extends 
to $900h^{-1}Mpc$, instead of 
sampling a complete sphere, the total sample 
 within the small solid angle was used and 
   $N(L,\Omega)\,=\,(\Omega/4\pi)BL^{D}$ was used to extract $D$. 
Now it is clear that this relation  is valid only for angular homogeneity.
The evidence of factal within $150h^{-1}Mpc$ came from LEDA, which was claimed
not to be  suitable 
for statistical analysis \cite{Davis}. Eliminating these from the 
results of analyses by P \cite{Pietronero},  the conclusion 
is that  there is no eveidence for fractal on scales  larger than 
$50h^{-1}Mpc$. However, if there is local  angular  fractal, 
which is also angularly-isotropic,
then $N(L,\Omega)\,\propto\,L^{D}$ is still 
valid though the rescaling is defferent in matching the results
 to  data obtained by 
using full conditional density. So there can be evidence of fractal 
for galaxy distribution on  
scales above $50h^{-1}Mpc$ 
only if it is independently proved to be isotropic.

Dramatically, there is such evidence, which just came from what was
 thought to be for
 homogeneity \cite{Davis}, such as isotropic distributions of
 microwave background radiation and other
 radiation, and 
 that in LCRS the distributions of number density as a function 
 of redshift is the same in six separate slices.
 In our opinion, actually they refer to 
angular-isotropy rather than homogeneity. Angular-isotropy may be satisfied by
  local angular 
  fractal or angular homogeneity. Even if angular homogeneity
  (in distribution) is proved, there is still the possibility that it is
  radially fractal, however, as discussed above, this model 
  has total dimension
  larger than 2 and put us in a privileged position  in the universe.
Given angular-isotropy, the total fractal dimension can be measured using numbers 
of points in a finite solid angle.

The dimension of local angular fractal can be measured by the dependence
of number on the angle. The radial distribution can be 
measured independently by pencil beam surveys, which was incorrectly 
interpreted by P as an intersection of $1$ dimension with $D$ dimension
in a $3d$ substrate, thus possess a minus dimension \cite{CP}.
 Actually it gives a subset of $D$ dimension, or say,
an intersection of $D_{r}$ dimension with $D$ dimension in the $D$ 
dimensional substrate. Even if the reported periodicity of $128h^{-1}Mpc$ 
\cite{Broadhurst} is confirmed to be due to the homogeneity of
radial distribution on corresponding scales, 
there can still be angular fractal.
 
 Before ending this letter, let us see what we can say about the
 controversy whether the boundness of cosmic fractal has been within
 observation \cite{Davis} \cite{Pietronero}. There has been much
 evidence for angular-isotropy. As exposed in this letter, this tells us the universe
 should be homogeneous or fractal with local angular fractal. 
 The evidence for
 homogeneity came from the IRAS redshift survey. Since the IRAS has a certain 
 degree of dilution, it was argued that the homogeneity behavior arises from
 diluteness \cite{Pietronero}, while the opposite argument stated that 
 diluteness cannot change the picture \cite{Davis}. We think that it depends 
 on how to dilute. If the diluteness is totally random with percentage
 $1-p$, then a fractal is still
 a fractal after dilution, since the number of points within a scale $r$
 only changes from $N(r)$ to $pN(r)$, without changing the fractal scaling.
 If the diluteness is related the distribution, a fractal can be 
 changed toward homogeneity. Another evidence for homogeneity is 
the unifom distribution 
 of $Ly-\alpha$ clouds \cite{Lu}\cite{Webb}\cite{Davis}.
 We think this only suggests the possibility of homogeneous distribution
 of intergalactic matter, without conflicting a fractal model of galaxy
 distribution. Is the galaxy distribution isotropic, rather than only
 angularly-isotropic on the correpsonding scales? Most likely yes, since
 from the observation \cite{Pietronero}\cite{CP} it is obtained that 
 $D\,\approx\,2$ while $D_{\Omega}\,\approx\,1.3$, thus $D_{r}\,\approx\,
 D_{\Omega}/2$.  
   
  In summary, the local angular fractal is claimed to cause the bahavior
  of angular projection, in  the way similar to that the behovior of two-point 
  correlation function is caused by the fractal in the whole space. A
  fractal with local angular fractal is angularly-isotropic. 
 As possessing a regularity between  homogeneity  and an ordinary fractal,
and a reconciliation of evidence claimed to be of homogeneity 
and that of fractal,
the local angular fractal may  be adopted by Nature in distributing
 galaxies. Further it is most likely to be isotropic revealed by
 observation. Whether the fractal observed in the large scale
  structure of the universe crossovers to homogeneity on a scale much smaller
  than the present observational limit deserves further investigation.
   
I. Kanter is thanked for hospitality in BIU where this work is done.

\newpage
Figure Caption:

Fig. 1. (a) $\omega(\theta)\,=\,(D_{\Omega}/2)
(\theta/\theta_{M})^{D_{\Omega}-2}-1$, and (b) $\Gamma(\theta)\,=\,
(AD_{\Omega}/2\pi)L^{D}\theta^{D_{\Omega}-2}$, setting $A\,=\,1$,
$D\,=\,2$, $D_{\Omega}\,=\,1.3$, $\theta_{M}\,=\,3.6^{\circ}\,=\,0.02\pi$
 and $L\,=\,20h^{-1}Mpc$. By comparing (a) and (b), it is clear 
 that $\theta^{*}$ where 
 $\omega(\theta^{*})\,=\,1$ and the deviation from
  power-law begins is spurious.
\newpage 
\begin{figure}
\setlength{\unitlength}{0.240900pt}
\ifx\plotpoint\undefined\newsavebox{\plotpoint}\fi
\begin{picture}(1500,900)(0,0)
\font\gnuplot=cmr10 at 10pt
\gnuplot
\sbox{\plotpoint}{\rule[-0.200pt]{0.400pt}{0.400pt}}%
\put(220.0,113.0){\rule[-0.200pt]{4.818pt}{0.400pt}}
\put(198,113){\makebox(0,0)[r]{0.01}}
\put(1416.0,113.0){\rule[-0.200pt]{4.818pt}{0.400pt}}
\put(220.0,170.0){\rule[-0.200pt]{2.409pt}{0.400pt}}
\put(1426.0,170.0){\rule[-0.200pt]{2.409pt}{0.400pt}}
\put(220.0,204.0){\rule[-0.200pt]{2.409pt}{0.400pt}}
\put(1426.0,204.0){\rule[-0.200pt]{2.409pt}{0.400pt}}
\put(220.0,228.0){\rule[-0.200pt]{2.409pt}{0.400pt}}
\put(1426.0,228.0){\rule[-0.200pt]{2.409pt}{0.400pt}}
\put(220.0,247.0){\rule[-0.200pt]{2.409pt}{0.400pt}}
\put(1426.0,247.0){\rule[-0.200pt]{2.409pt}{0.400pt}}
\put(220.0,262.0){\rule[-0.200pt]{2.409pt}{0.400pt}}
\put(1426.0,262.0){\rule[-0.200pt]{2.409pt}{0.400pt}}
\put(220.0,274.0){\rule[-0.200pt]{2.409pt}{0.400pt}}
\put(1426.0,274.0){\rule[-0.200pt]{2.409pt}{0.400pt}}
\put(220.0,285.0){\rule[-0.200pt]{2.409pt}{0.400pt}}
\put(1426.0,285.0){\rule[-0.200pt]{2.409pt}{0.400pt}}
\put(220.0,295.0){\rule[-0.200pt]{2.409pt}{0.400pt}}
\put(1426.0,295.0){\rule[-0.200pt]{2.409pt}{0.400pt}}
\put(220.0,304.0){\rule[-0.200pt]{4.818pt}{0.400pt}}
\put(198,304){\makebox(0,0)[r]{0.1}}
\put(1416.0,304.0){\rule[-0.200pt]{4.818pt}{0.400pt}}
\put(220.0,361.0){\rule[-0.200pt]{2.409pt}{0.400pt}}
\put(1426.0,361.0){\rule[-0.200pt]{2.409pt}{0.400pt}}
\put(220.0,395.0){\rule[-0.200pt]{2.409pt}{0.400pt}}
\put(1426.0,395.0){\rule[-0.200pt]{2.409pt}{0.400pt}}
\put(220.0,419.0){\rule[-0.200pt]{2.409pt}{0.400pt}}
\put(1426.0,419.0){\rule[-0.200pt]{2.409pt}{0.400pt}}
\put(220.0,438.0){\rule[-0.200pt]{2.409pt}{0.400pt}}
\put(1426.0,438.0){\rule[-0.200pt]{2.409pt}{0.400pt}}
\put(220.0,453.0){\rule[-0.200pt]{2.409pt}{0.400pt}}
\put(1426.0,453.0){\rule[-0.200pt]{2.409pt}{0.400pt}}
\put(220.0,465.0){\rule[-0.200pt]{2.409pt}{0.400pt}}
\put(1426.0,465.0){\rule[-0.200pt]{2.409pt}{0.400pt}}
\put(220.0,476.0){\rule[-0.200pt]{2.409pt}{0.400pt}}
\put(1426.0,476.0){\rule[-0.200pt]{2.409pt}{0.400pt}}
\put(220.0,486.0){\rule[-0.200pt]{2.409pt}{0.400pt}}
\put(1426.0,486.0){\rule[-0.200pt]{2.409pt}{0.400pt}}
\put(220.0,495.0){\rule[-0.200pt]{4.818pt}{0.400pt}}
\put(198,495){\makebox(0,0)[r]{1}}
\put(1416.0,495.0){\rule[-0.200pt]{4.818pt}{0.400pt}}
\put(220.0,552.0){\rule[-0.200pt]{2.409pt}{0.400pt}}
\put(1426.0,552.0){\rule[-0.200pt]{2.409pt}{0.400pt}}
\put(220.0,586.0){\rule[-0.200pt]{2.409pt}{0.400pt}}
\put(1426.0,586.0){\rule[-0.200pt]{2.409pt}{0.400pt}}
\put(220.0,610.0){\rule[-0.200pt]{2.409pt}{0.400pt}}
\put(1426.0,610.0){\rule[-0.200pt]{2.409pt}{0.400pt}}
\put(220.0,629.0){\rule[-0.200pt]{2.409pt}{0.400pt}}
\put(1426.0,629.0){\rule[-0.200pt]{2.409pt}{0.400pt}}
\put(220.0,644.0){\rule[-0.200pt]{2.409pt}{0.400pt}}
\put(1426.0,644.0){\rule[-0.200pt]{2.409pt}{0.400pt}}
\put(220.0,656.0){\rule[-0.200pt]{2.409pt}{0.400pt}}
\put(1426.0,656.0){\rule[-0.200pt]{2.409pt}{0.400pt}}
\put(220.0,667.0){\rule[-0.200pt]{2.409pt}{0.400pt}}
\put(1426.0,667.0){\rule[-0.200pt]{2.409pt}{0.400pt}}
\put(220.0,677.0){\rule[-0.200pt]{2.409pt}{0.400pt}}
\put(1426.0,677.0){\rule[-0.200pt]{2.409pt}{0.400pt}}
\put(220.0,686.0){\rule[-0.200pt]{4.818pt}{0.400pt}}
\put(198,686){\makebox(0,0)[r]{10}}
\put(1416.0,686.0){\rule[-0.200pt]{4.818pt}{0.400pt}}
\put(220.0,743.0){\rule[-0.200pt]{2.409pt}{0.400pt}}
\put(1426.0,743.0){\rule[-0.200pt]{2.409pt}{0.400pt}}
\put(220.0,777.0){\rule[-0.200pt]{2.409pt}{0.400pt}}
\put(1426.0,777.0){\rule[-0.200pt]{2.409pt}{0.400pt}}
\put(220.0,801.0){\rule[-0.200pt]{2.409pt}{0.400pt}}
\put(1426.0,801.0){\rule[-0.200pt]{2.409pt}{0.400pt}}
\put(220.0,820.0){\rule[-0.200pt]{2.409pt}{0.400pt}}
\put(1426.0,820.0){\rule[-0.200pt]{2.409pt}{0.400pt}}
\put(220.0,835.0){\rule[-0.200pt]{2.409pt}{0.400pt}}
\put(1426.0,835.0){\rule[-0.200pt]{2.409pt}{0.400pt}}
\put(220.0,847.0){\rule[-0.200pt]{2.409pt}{0.400pt}}
\put(1426.0,847.0){\rule[-0.200pt]{2.409pt}{0.400pt}}
\put(220.0,858.0){\rule[-0.200pt]{2.409pt}{0.400pt}}
\put(1426.0,858.0){\rule[-0.200pt]{2.409pt}{0.400pt}}
\put(220.0,868.0){\rule[-0.200pt]{2.409pt}{0.400pt}}
\put(1426.0,868.0){\rule[-0.200pt]{2.409pt}{0.400pt}}
\put(220.0,877.0){\rule[-0.200pt]{4.818pt}{0.400pt}}
\put(198,877){\makebox(0,0)[r]{100}}
\put(1416.0,877.0){\rule[-0.200pt]{4.818pt}{0.400pt}}
\put(220.0,113.0){\rule[-0.200pt]{0.400pt}{4.818pt}}
\put(220,68){\makebox(0,0){0.001}}
\put(220.0,857.0){\rule[-0.200pt]{0.400pt}{4.818pt}}
\put(424.0,113.0){\rule[-0.200pt]{0.400pt}{2.409pt}}
\put(424.0,867.0){\rule[-0.200pt]{0.400pt}{2.409pt}}
\put(543.0,113.0){\rule[-0.200pt]{0.400pt}{2.409pt}}
\put(543.0,867.0){\rule[-0.200pt]{0.400pt}{2.409pt}}
\put(627.0,113.0){\rule[-0.200pt]{0.400pt}{2.409pt}}
\put(627.0,867.0){\rule[-0.200pt]{0.400pt}{2.409pt}}
\put(693.0,113.0){\rule[-0.200pt]{0.400pt}{2.409pt}}
\put(693.0,867.0){\rule[-0.200pt]{0.400pt}{2.409pt}}
\put(746.0,113.0){\rule[-0.200pt]{0.400pt}{2.409pt}}
\put(746.0,867.0){\rule[-0.200pt]{0.400pt}{2.409pt}}
\put(792.0,113.0){\rule[-0.200pt]{0.400pt}{2.409pt}}
\put(792.0,867.0){\rule[-0.200pt]{0.400pt}{2.409pt}}
\put(831.0,113.0){\rule[-0.200pt]{0.400pt}{2.409pt}}
\put(831.0,867.0){\rule[-0.200pt]{0.400pt}{2.409pt}}
\put(865.0,113.0){\rule[-0.200pt]{0.400pt}{2.409pt}}
\put(865.0,867.0){\rule[-0.200pt]{0.400pt}{2.409pt}}
\put(896.0,113.0){\rule[-0.200pt]{0.400pt}{4.818pt}}
\put(896,68){\makebox(0,0){0.01}}
\put(896.0,857.0){\rule[-0.200pt]{0.400pt}{4.818pt}}
\put(1100.0,113.0){\rule[-0.200pt]{0.400pt}{2.409pt}}
\put(1100.0,867.0){\rule[-0.200pt]{0.400pt}{2.409pt}}
\put(1219.0,113.0){\rule[-0.200pt]{0.400pt}{2.409pt}}
\put(1219.0,867.0){\rule[-0.200pt]{0.400pt}{2.409pt}}
\put(1304.0,113.0){\rule[-0.200pt]{0.400pt}{2.409pt}}
\put(1304.0,867.0){\rule[-0.200pt]{0.400pt}{2.409pt}}
\put(1369.0,113.0){\rule[-0.200pt]{0.400pt}{2.409pt}}
\put(1369.0,867.0){\rule[-0.200pt]{0.400pt}{2.409pt}}
\put(1423.0,113.0){\rule[-0.200pt]{0.400pt}{2.409pt}}
\put(1423.0,867.0){\rule[-0.200pt]{0.400pt}{2.409pt}}
\put(220.0,113.0){\rule[-0.200pt]{292.934pt}{0.400pt}}
\put(1436.0,113.0){\rule[-0.200pt]{0.400pt}{184.048pt}}
\put(220.0,877.0){\rule[-0.200pt]{292.934pt}{0.400pt}}
\put(45,495){\makebox(0,0){$\omega(\theta)$}}
\put(828,23){\makebox(0,0){$\theta$}}
\put(896,801){\makebox(0,0){(a)}}
\put(220.0,113.0){\rule[-0.200pt]{0.400pt}{184.048pt}}
\put(220,616){\usebox{\plotpoint}}
\multiput(220.00,614.95)(2.472,-0.447){3}{\rule{1.700pt}{0.108pt}}
\multiput(220.00,615.17)(8.472,-3.000){2}{\rule{0.850pt}{0.400pt}}
\multiput(232.00,611.95)(2.695,-0.447){3}{\rule{1.833pt}{0.108pt}}
\multiput(232.00,612.17)(9.195,-3.000){2}{\rule{0.917pt}{0.400pt}}
\multiput(245.00,608.95)(2.472,-0.447){3}{\rule{1.700pt}{0.108pt}}
\multiput(245.00,609.17)(8.472,-3.000){2}{\rule{0.850pt}{0.400pt}}
\multiput(257.00,605.95)(2.472,-0.447){3}{\rule{1.700pt}{0.108pt}}
\multiput(257.00,606.17)(8.472,-3.000){2}{\rule{0.850pt}{0.400pt}}
\multiput(269.00,602.95)(2.472,-0.447){3}{\rule{1.700pt}{0.108pt}}
\multiput(269.00,603.17)(8.472,-3.000){2}{\rule{0.850pt}{0.400pt}}
\multiput(281.00,599.94)(1.797,-0.468){5}{\rule{1.400pt}{0.113pt}}
\multiput(281.00,600.17)(10.094,-4.000){2}{\rule{0.700pt}{0.400pt}}
\multiput(294.00,595.95)(2.472,-0.447){3}{\rule{1.700pt}{0.108pt}}
\multiput(294.00,596.17)(8.472,-3.000){2}{\rule{0.850pt}{0.400pt}}
\multiput(306.00,592.95)(2.472,-0.447){3}{\rule{1.700pt}{0.108pt}}
\multiput(306.00,593.17)(8.472,-3.000){2}{\rule{0.850pt}{0.400pt}}
\multiput(318.00,589.95)(2.695,-0.447){3}{\rule{1.833pt}{0.108pt}}
\multiput(318.00,590.17)(9.195,-3.000){2}{\rule{0.917pt}{0.400pt}}
\multiput(331.00,586.95)(2.472,-0.447){3}{\rule{1.700pt}{0.108pt}}
\multiput(331.00,587.17)(8.472,-3.000){2}{\rule{0.850pt}{0.400pt}}
\multiput(343.00,583.94)(1.651,-0.468){5}{\rule{1.300pt}{0.113pt}}
\multiput(343.00,584.17)(9.302,-4.000){2}{\rule{0.650pt}{0.400pt}}
\multiput(355.00,579.95)(2.472,-0.447){3}{\rule{1.700pt}{0.108pt}}
\multiput(355.00,580.17)(8.472,-3.000){2}{\rule{0.850pt}{0.400pt}}
\multiput(367.00,576.95)(2.695,-0.447){3}{\rule{1.833pt}{0.108pt}}
\multiput(367.00,577.17)(9.195,-3.000){2}{\rule{0.917pt}{0.400pt}}
\multiput(380.00,573.94)(1.651,-0.468){5}{\rule{1.300pt}{0.113pt}}
\multiput(380.00,574.17)(9.302,-4.000){2}{\rule{0.650pt}{0.400pt}}
\multiput(392.00,569.95)(2.472,-0.447){3}{\rule{1.700pt}{0.108pt}}
\multiput(392.00,570.17)(8.472,-3.000){2}{\rule{0.850pt}{0.400pt}}
\multiput(404.00,566.95)(2.695,-0.447){3}{\rule{1.833pt}{0.108pt}}
\multiput(404.00,567.17)(9.195,-3.000){2}{\rule{0.917pt}{0.400pt}}
\multiput(417.00,563.94)(1.651,-0.468){5}{\rule{1.300pt}{0.113pt}}
\multiput(417.00,564.17)(9.302,-4.000){2}{\rule{0.650pt}{0.400pt}}
\multiput(429.00,559.95)(2.472,-0.447){3}{\rule{1.700pt}{0.108pt}}
\multiput(429.00,560.17)(8.472,-3.000){2}{\rule{0.850pt}{0.400pt}}
\multiput(441.00,556.94)(1.651,-0.468){5}{\rule{1.300pt}{0.113pt}}
\multiput(441.00,557.17)(9.302,-4.000){2}{\rule{0.650pt}{0.400pt}}
\multiput(453.00,552.94)(1.797,-0.468){5}{\rule{1.400pt}{0.113pt}}
\multiput(453.00,553.17)(10.094,-4.000){2}{\rule{0.700pt}{0.400pt}}
\multiput(466.00,548.95)(2.472,-0.447){3}{\rule{1.700pt}{0.108pt}}
\multiput(466.00,549.17)(8.472,-3.000){2}{\rule{0.850pt}{0.400pt}}
\multiput(478.00,545.94)(1.651,-0.468){5}{\rule{1.300pt}{0.113pt}}
\multiput(478.00,546.17)(9.302,-4.000){2}{\rule{0.650pt}{0.400pt}}
\multiput(490.00,541.94)(1.797,-0.468){5}{\rule{1.400pt}{0.113pt}}
\multiput(490.00,542.17)(10.094,-4.000){2}{\rule{0.700pt}{0.400pt}}
\multiput(503.00,537.94)(1.651,-0.468){5}{\rule{1.300pt}{0.113pt}}
\multiput(503.00,538.17)(9.302,-4.000){2}{\rule{0.650pt}{0.400pt}}
\multiput(515.00,533.94)(1.651,-0.468){5}{\rule{1.300pt}{0.113pt}}
\multiput(515.00,534.17)(9.302,-4.000){2}{\rule{0.650pt}{0.400pt}}
\multiput(527.00,529.94)(1.651,-0.468){5}{\rule{1.300pt}{0.113pt}}
\multiput(527.00,530.17)(9.302,-4.000){2}{\rule{0.650pt}{0.400pt}}
\multiput(539.00,525.94)(1.797,-0.468){5}{\rule{1.400pt}{0.113pt}}
\multiput(539.00,526.17)(10.094,-4.000){2}{\rule{0.700pt}{0.400pt}}
\multiput(552.00,521.94)(1.651,-0.468){5}{\rule{1.300pt}{0.113pt}}
\multiput(552.00,522.17)(9.302,-4.000){2}{\rule{0.650pt}{0.400pt}}
\multiput(564.00,517.93)(1.267,-0.477){7}{\rule{1.060pt}{0.115pt}}
\multiput(564.00,518.17)(9.800,-5.000){2}{\rule{0.530pt}{0.400pt}}
\multiput(576.00,512.94)(1.651,-0.468){5}{\rule{1.300pt}{0.113pt}}
\multiput(576.00,513.17)(9.302,-4.000){2}{\rule{0.650pt}{0.400pt}}
\multiput(588.00,508.94)(1.797,-0.468){5}{\rule{1.400pt}{0.113pt}}
\multiput(588.00,509.17)(10.094,-4.000){2}{\rule{0.700pt}{0.400pt}}
\multiput(601.00,504.93)(1.267,-0.477){7}{\rule{1.060pt}{0.115pt}}
\multiput(601.00,505.17)(9.800,-5.000){2}{\rule{0.530pt}{0.400pt}}
\multiput(613.00,499.93)(1.267,-0.477){7}{\rule{1.060pt}{0.115pt}}
\multiput(613.00,500.17)(9.800,-5.000){2}{\rule{0.530pt}{0.400pt}}
\multiput(625.00,494.93)(1.378,-0.477){7}{\rule{1.140pt}{0.115pt}}
\multiput(625.00,495.17)(10.634,-5.000){2}{\rule{0.570pt}{0.400pt}}
\multiput(638.00,489.93)(1.267,-0.477){7}{\rule{1.060pt}{0.115pt}}
\multiput(638.00,490.17)(9.800,-5.000){2}{\rule{0.530pt}{0.400pt}}
\multiput(650.00,484.93)(1.267,-0.477){7}{\rule{1.060pt}{0.115pt}}
\multiput(650.00,485.17)(9.800,-5.000){2}{\rule{0.530pt}{0.400pt}}
\multiput(662.00,479.93)(1.267,-0.477){7}{\rule{1.060pt}{0.115pt}}
\multiput(662.00,480.17)(9.800,-5.000){2}{\rule{0.530pt}{0.400pt}}
\multiput(674.00,474.93)(1.123,-0.482){9}{\rule{0.967pt}{0.116pt}}
\multiput(674.00,475.17)(10.994,-6.000){2}{\rule{0.483pt}{0.400pt}}
\multiput(687.00,468.93)(1.033,-0.482){9}{\rule{0.900pt}{0.116pt}}
\multiput(687.00,469.17)(10.132,-6.000){2}{\rule{0.450pt}{0.400pt}}
\multiput(699.00,462.93)(1.033,-0.482){9}{\rule{0.900pt}{0.116pt}}
\multiput(699.00,463.17)(10.132,-6.000){2}{\rule{0.450pt}{0.400pt}}
\multiput(711.00,456.93)(1.123,-0.482){9}{\rule{0.967pt}{0.116pt}}
\multiput(711.00,457.17)(10.994,-6.000){2}{\rule{0.483pt}{0.400pt}}
\multiput(724.00,450.93)(0.874,-0.485){11}{\rule{0.786pt}{0.117pt}}
\multiput(724.00,451.17)(10.369,-7.000){2}{\rule{0.393pt}{0.400pt}}
\multiput(736.00,443.93)(0.874,-0.485){11}{\rule{0.786pt}{0.117pt}}
\multiput(736.00,444.17)(10.369,-7.000){2}{\rule{0.393pt}{0.400pt}}
\multiput(748.00,436.93)(0.874,-0.485){11}{\rule{0.786pt}{0.117pt}}
\multiput(748.00,437.17)(10.369,-7.000){2}{\rule{0.393pt}{0.400pt}}
\multiput(760.00,429.93)(0.824,-0.488){13}{\rule{0.750pt}{0.117pt}}
\multiput(760.00,430.17)(11.443,-8.000){2}{\rule{0.375pt}{0.400pt}}
\multiput(773.00,421.93)(0.669,-0.489){15}{\rule{0.633pt}{0.118pt}}
\multiput(773.00,422.17)(10.685,-9.000){2}{\rule{0.317pt}{0.400pt}}
\multiput(785.00,412.93)(0.669,-0.489){15}{\rule{0.633pt}{0.118pt}}
\multiput(785.00,413.17)(10.685,-9.000){2}{\rule{0.317pt}{0.400pt}}
\multiput(797.00,403.92)(0.652,-0.491){17}{\rule{0.620pt}{0.118pt}}
\multiput(797.00,404.17)(11.713,-10.000){2}{\rule{0.310pt}{0.400pt}}
\multiput(810.00,393.92)(0.543,-0.492){19}{\rule{0.536pt}{0.118pt}}
\multiput(810.00,394.17)(10.887,-11.000){2}{\rule{0.268pt}{0.400pt}}
\multiput(822.58,381.79)(0.492,-0.539){21}{\rule{0.119pt}{0.533pt}}
\multiput(821.17,382.89)(12.000,-11.893){2}{\rule{0.400pt}{0.267pt}}
\multiput(834.58,368.65)(0.492,-0.582){21}{\rule{0.119pt}{0.567pt}}
\multiput(833.17,369.82)(12.000,-12.824){2}{\rule{0.400pt}{0.283pt}}
\multiput(846.58,354.41)(0.493,-0.655){23}{\rule{0.119pt}{0.623pt}}
\multiput(845.17,355.71)(13.000,-15.707){2}{\rule{0.400pt}{0.312pt}}
\multiput(859.58,336.82)(0.492,-0.841){21}{\rule{0.119pt}{0.767pt}}
\multiput(858.17,338.41)(12.000,-18.409){2}{\rule{0.400pt}{0.383pt}}
\multiput(871.58,316.13)(0.492,-1.056){21}{\rule{0.119pt}{0.933pt}}
\multiput(870.17,318.06)(12.000,-23.063){2}{\rule{0.400pt}{0.467pt}}
\multiput(883.58,289.99)(0.493,-1.408){23}{\rule{0.119pt}{1.208pt}}
\multiput(882.17,292.49)(13.000,-33.493){2}{\rule{0.400pt}{0.604pt}}
\multiput(896.58,249.87)(0.492,-2.693){21}{\rule{0.119pt}{2.200pt}}
\multiput(895.17,254.43)(12.000,-58.434){2}{\rule{0.400pt}{1.100pt}}
\end{picture}
\vspace{1cm}
\setlength{\unitlength}{0.240900pt}
\ifx\plotpoint\undefined\newsavebox{\plotpoint}\fi
\begin{picture}(1500,900)(0,0)
\font\gnuplot=cmr10 at 10pt
\gnuplot
\sbox{\plotpoint}{\rule[-0.200pt]{0.400pt}{0.400pt}}%
\put(220.0,113.0){\rule[-0.200pt]{4.818pt}{0.400pt}}
\put(198,113){\makebox(0,0)[r]{10}}
\put(1416.0,113.0){\rule[-0.200pt]{4.818pt}{0.400pt}}
\put(220.0,170.0){\rule[-0.200pt]{2.409pt}{0.400pt}}
\put(1426.0,170.0){\rule[-0.200pt]{2.409pt}{0.400pt}}
\put(220.0,204.0){\rule[-0.200pt]{2.409pt}{0.400pt}}
\put(1426.0,204.0){\rule[-0.200pt]{2.409pt}{0.400pt}}
\put(220.0,228.0){\rule[-0.200pt]{2.409pt}{0.400pt}}
\put(1426.0,228.0){\rule[-0.200pt]{2.409pt}{0.400pt}}
\put(220.0,247.0){\rule[-0.200pt]{2.409pt}{0.400pt}}
\put(1426.0,247.0){\rule[-0.200pt]{2.409pt}{0.400pt}}
\put(220.0,262.0){\rule[-0.200pt]{2.409pt}{0.400pt}}
\put(1426.0,262.0){\rule[-0.200pt]{2.409pt}{0.400pt}}
\put(220.0,274.0){\rule[-0.200pt]{2.409pt}{0.400pt}}
\put(1426.0,274.0){\rule[-0.200pt]{2.409pt}{0.400pt}}
\put(220.0,285.0){\rule[-0.200pt]{2.409pt}{0.400pt}}
\put(1426.0,285.0){\rule[-0.200pt]{2.409pt}{0.400pt}}
\put(220.0,295.0){\rule[-0.200pt]{2.409pt}{0.400pt}}
\put(1426.0,295.0){\rule[-0.200pt]{2.409pt}{0.400pt}}
\put(220.0,304.0){\rule[-0.200pt]{4.818pt}{0.400pt}}
\put(198,304){\makebox(0,0)[r]{100}}
\put(1416.0,304.0){\rule[-0.200pt]{4.818pt}{0.400pt}}
\put(220.0,361.0){\rule[-0.200pt]{2.409pt}{0.400pt}}
\put(1426.0,361.0){\rule[-0.200pt]{2.409pt}{0.400pt}}
\put(220.0,395.0){\rule[-0.200pt]{2.409pt}{0.400pt}}
\put(1426.0,395.0){\rule[-0.200pt]{2.409pt}{0.400pt}}
\put(220.0,419.0){\rule[-0.200pt]{2.409pt}{0.400pt}}
\put(1426.0,419.0){\rule[-0.200pt]{2.409pt}{0.400pt}}
\put(220.0,438.0){\rule[-0.200pt]{2.409pt}{0.400pt}}
\put(1426.0,438.0){\rule[-0.200pt]{2.409pt}{0.400pt}}
\put(220.0,453.0){\rule[-0.200pt]{2.409pt}{0.400pt}}
\put(1426.0,453.0){\rule[-0.200pt]{2.409pt}{0.400pt}}
\put(220.0,465.0){\rule[-0.200pt]{2.409pt}{0.400pt}}
\put(1426.0,465.0){\rule[-0.200pt]{2.409pt}{0.400pt}}
\put(220.0,476.0){\rule[-0.200pt]{2.409pt}{0.400pt}}
\put(1426.0,476.0){\rule[-0.200pt]{2.409pt}{0.400pt}}
\put(220.0,486.0){\rule[-0.200pt]{2.409pt}{0.400pt}}
\put(1426.0,486.0){\rule[-0.200pt]{2.409pt}{0.400pt}}
\put(220.0,495.0){\rule[-0.200pt]{4.818pt}{0.400pt}}
\put(198,495){\makebox(0,0)[r]{1000}}
\put(1416.0,495.0){\rule[-0.200pt]{4.818pt}{0.400pt}}
\put(220.0,552.0){\rule[-0.200pt]{2.409pt}{0.400pt}}
\put(1426.0,552.0){\rule[-0.200pt]{2.409pt}{0.400pt}}
\put(220.0,586.0){\rule[-0.200pt]{2.409pt}{0.400pt}}
\put(1426.0,586.0){\rule[-0.200pt]{2.409pt}{0.400pt}}
\put(220.0,610.0){\rule[-0.200pt]{2.409pt}{0.400pt}}
\put(1426.0,610.0){\rule[-0.200pt]{2.409pt}{0.400pt}}
\put(220.0,629.0){\rule[-0.200pt]{2.409pt}{0.400pt}}
\put(1426.0,629.0){\rule[-0.200pt]{2.409pt}{0.400pt}}
\put(220.0,644.0){\rule[-0.200pt]{2.409pt}{0.400pt}}
\put(1426.0,644.0){\rule[-0.200pt]{2.409pt}{0.400pt}}
\put(220.0,656.0){\rule[-0.200pt]{2.409pt}{0.400pt}}
\put(1426.0,656.0){\rule[-0.200pt]{2.409pt}{0.400pt}}
\put(220.0,667.0){\rule[-0.200pt]{2.409pt}{0.400pt}}
\put(1426.0,667.0){\rule[-0.200pt]{2.409pt}{0.400pt}}
\put(220.0,677.0){\rule[-0.200pt]{2.409pt}{0.400pt}}
\put(1426.0,677.0){\rule[-0.200pt]{2.409pt}{0.400pt}}
\put(220.0,686.0){\rule[-0.200pt]{4.818pt}{0.400pt}}
\put(198,686){\makebox(0,0)[r]{10000}}
\put(1416.0,686.0){\rule[-0.200pt]{4.818pt}{0.400pt}}
\put(220.0,743.0){\rule[-0.200pt]{2.409pt}{0.400pt}}
\put(1426.0,743.0){\rule[-0.200pt]{2.409pt}{0.400pt}}
\put(220.0,777.0){\rule[-0.200pt]{2.409pt}{0.400pt}}
\put(1426.0,777.0){\rule[-0.200pt]{2.409pt}{0.400pt}}
\put(220.0,801.0){\rule[-0.200pt]{2.409pt}{0.400pt}}
\put(1426.0,801.0){\rule[-0.200pt]{2.409pt}{0.400pt}}
\put(220.0,820.0){\rule[-0.200pt]{2.409pt}{0.400pt}}
\put(1426.0,820.0){\rule[-0.200pt]{2.409pt}{0.400pt}}
\put(220.0,835.0){\rule[-0.200pt]{2.409pt}{0.400pt}}
\put(1426.0,835.0){\rule[-0.200pt]{2.409pt}{0.400pt}}
\put(220.0,847.0){\rule[-0.200pt]{2.409pt}{0.400pt}}
\put(1426.0,847.0){\rule[-0.200pt]{2.409pt}{0.400pt}}
\put(220.0,858.0){\rule[-0.200pt]{2.409pt}{0.400pt}}
\put(1426.0,858.0){\rule[-0.200pt]{2.409pt}{0.400pt}}
\put(220.0,868.0){\rule[-0.200pt]{2.409pt}{0.400pt}}
\put(1426.0,868.0){\rule[-0.200pt]{2.409pt}{0.400pt}}
\put(220.0,877.0){\rule[-0.200pt]{4.818pt}{0.400pt}}
\put(198,877){\makebox(0,0)[r]{100000}}
\put(1416.0,877.0){\rule[-0.200pt]{4.818pt}{0.400pt}}
\put(220.0,113.0){\rule[-0.200pt]{0.400pt}{4.818pt}}
\put(220,68){\makebox(0,0){0.001}}
\put(220.0,857.0){\rule[-0.200pt]{0.400pt}{4.818pt}}
\put(424.0,113.0){\rule[-0.200pt]{0.400pt}{2.409pt}}
\put(424.0,867.0){\rule[-0.200pt]{0.400pt}{2.409pt}}
\put(543.0,113.0){\rule[-0.200pt]{0.400pt}{2.409pt}}
\put(543.0,867.0){\rule[-0.200pt]{0.400pt}{2.409pt}}
\put(627.0,113.0){\rule[-0.200pt]{0.400pt}{2.409pt}}
\put(627.0,867.0){\rule[-0.200pt]{0.400pt}{2.409pt}}
\put(693.0,113.0){\rule[-0.200pt]{0.400pt}{2.409pt}}
\put(693.0,867.0){\rule[-0.200pt]{0.400pt}{2.409pt}}
\put(746.0,113.0){\rule[-0.200pt]{0.400pt}{2.409pt}}
\put(746.0,867.0){\rule[-0.200pt]{0.400pt}{2.409pt}}
\put(792.0,113.0){\rule[-0.200pt]{0.400pt}{2.409pt}}
\put(792.0,867.0){\rule[-0.200pt]{0.400pt}{2.409pt}}
\put(831.0,113.0){\rule[-0.200pt]{0.400pt}{2.409pt}}
\put(831.0,867.0){\rule[-0.200pt]{0.400pt}{2.409pt}}
\put(865.0,113.0){\rule[-0.200pt]{0.400pt}{2.409pt}}
\put(865.0,867.0){\rule[-0.200pt]{0.400pt}{2.409pt}}
\put(896.0,113.0){\rule[-0.200pt]{0.400pt}{4.818pt}}
\put(896,68){\makebox(0,0){0.01}}
\put(896.0,857.0){\rule[-0.200pt]{0.400pt}{4.818pt}}
\put(1100.0,113.0){\rule[-0.200pt]{0.400pt}{2.409pt}}
\put(1100.0,867.0){\rule[-0.200pt]{0.400pt}{2.409pt}}
\put(1219.0,113.0){\rule[-0.200pt]{0.400pt}{2.409pt}}
\put(1219.0,867.0){\rule[-0.200pt]{0.400pt}{2.409pt}}
\put(1304.0,113.0){\rule[-0.200pt]{0.400pt}{2.409pt}}
\put(1304.0,867.0){\rule[-0.200pt]{0.400pt}{2.409pt}}
\put(1369.0,113.0){\rule[-0.200pt]{0.400pt}{2.409pt}}
\put(1369.0,867.0){\rule[-0.200pt]{0.400pt}{2.409pt}}
\put(1423.0,113.0){\rule[-0.200pt]{0.400pt}{2.409pt}}
\put(1423.0,867.0){\rule[-0.200pt]{0.400pt}{2.409pt}}
\put(220.0,113.0){\rule[-0.200pt]{292.934pt}{0.400pt}}
\put(1436.0,113.0){\rule[-0.200pt]{0.400pt}{184.048pt}}
\put(220.0,877.0){\rule[-0.200pt]{292.934pt}{0.400pt}}
\put(45,495){\makebox(0,0){$\Gamma(\theta)$}}
\put(828,23){\makebox(0,0){$\theta$}}
\put(896,801){\makebox(0,0){(b)}}
\put(220.0,113.0){\rule[-0.200pt]{0.400pt}{184.048pt}}
\put(220,689){\usebox{\plotpoint}}
\put(220,687.17){\rule{2.500pt}{0.400pt}}
\multiput(220.00,688.17)(6.811,-2.000){2}{\rule{1.250pt}{0.400pt}}
\put(232,685.17){\rule{2.700pt}{0.400pt}}
\multiput(232.00,686.17)(7.396,-2.000){2}{\rule{1.350pt}{0.400pt}}
\multiput(245.00,683.95)(2.472,-0.447){3}{\rule{1.700pt}{0.108pt}}
\multiput(245.00,684.17)(8.472,-3.000){2}{\rule{0.850pt}{0.400pt}}
\put(257,680.17){\rule{2.500pt}{0.400pt}}
\multiput(257.00,681.17)(6.811,-2.000){2}{\rule{1.250pt}{0.400pt}}
\multiput(269.00,678.95)(2.472,-0.447){3}{\rule{1.700pt}{0.108pt}}
\multiput(269.00,679.17)(8.472,-3.000){2}{\rule{0.850pt}{0.400pt}}
\put(281,675.17){\rule{2.700pt}{0.400pt}}
\multiput(281.00,676.17)(7.396,-2.000){2}{\rule{1.350pt}{0.400pt}}
\multiput(294.00,673.95)(2.472,-0.447){3}{\rule{1.700pt}{0.108pt}}
\multiput(294.00,674.17)(8.472,-3.000){2}{\rule{0.850pt}{0.400pt}}
\put(306,670.17){\rule{2.500pt}{0.400pt}}
\multiput(306.00,671.17)(6.811,-2.000){2}{\rule{1.250pt}{0.400pt}}
\put(318,668.17){\rule{2.700pt}{0.400pt}}
\multiput(318.00,669.17)(7.396,-2.000){2}{\rule{1.350pt}{0.400pt}}
\multiput(331.00,666.95)(2.472,-0.447){3}{\rule{1.700pt}{0.108pt}}
\multiput(331.00,667.17)(8.472,-3.000){2}{\rule{0.850pt}{0.400pt}}
\put(343,663.17){\rule{2.500pt}{0.400pt}}
\multiput(343.00,664.17)(6.811,-2.000){2}{\rule{1.250pt}{0.400pt}}
\multiput(355.00,661.95)(2.472,-0.447){3}{\rule{1.700pt}{0.108pt}}
\multiput(355.00,662.17)(8.472,-3.000){2}{\rule{0.850pt}{0.400pt}}
\put(367,658.17){\rule{2.700pt}{0.400pt}}
\multiput(367.00,659.17)(7.396,-2.000){2}{\rule{1.350pt}{0.400pt}}
\multiput(380.00,656.95)(2.472,-0.447){3}{\rule{1.700pt}{0.108pt}}
\multiput(380.00,657.17)(8.472,-3.000){2}{\rule{0.850pt}{0.400pt}}
\put(392,653.17){\rule{2.500pt}{0.400pt}}
\multiput(392.00,654.17)(6.811,-2.000){2}{\rule{1.250pt}{0.400pt}}
\put(404,651.17){\rule{2.700pt}{0.400pt}}
\multiput(404.00,652.17)(7.396,-2.000){2}{\rule{1.350pt}{0.400pt}}
\multiput(417.00,649.95)(2.472,-0.447){3}{\rule{1.700pt}{0.108pt}}
\multiput(417.00,650.17)(8.472,-3.000){2}{\rule{0.850pt}{0.400pt}}
\put(429,646.17){\rule{2.500pt}{0.400pt}}
\multiput(429.00,647.17)(6.811,-2.000){2}{\rule{1.250pt}{0.400pt}}
\multiput(441.00,644.95)(2.472,-0.447){3}{\rule{1.700pt}{0.108pt}}
\multiput(441.00,645.17)(8.472,-3.000){2}{\rule{0.850pt}{0.400pt}}
\put(453,641.17){\rule{2.700pt}{0.400pt}}
\multiput(453.00,642.17)(7.396,-2.000){2}{\rule{1.350pt}{0.400pt}}
\multiput(466.00,639.95)(2.472,-0.447){3}{\rule{1.700pt}{0.108pt}}
\multiput(466.00,640.17)(8.472,-3.000){2}{\rule{0.850pt}{0.400pt}}
\put(478,636.17){\rule{2.500pt}{0.400pt}}
\multiput(478.00,637.17)(6.811,-2.000){2}{\rule{1.250pt}{0.400pt}}
\put(490,634.17){\rule{2.700pt}{0.400pt}}
\multiput(490.00,635.17)(7.396,-2.000){2}{\rule{1.350pt}{0.400pt}}
\multiput(503.00,632.95)(2.472,-0.447){3}{\rule{1.700pt}{0.108pt}}
\multiput(503.00,633.17)(8.472,-3.000){2}{\rule{0.850pt}{0.400pt}}
\put(515,629.17){\rule{2.500pt}{0.400pt}}
\multiput(515.00,630.17)(6.811,-2.000){2}{\rule{1.250pt}{0.400pt}}
\multiput(527.00,627.95)(2.472,-0.447){3}{\rule{1.700pt}{0.108pt}}
\multiput(527.00,628.17)(8.472,-3.000){2}{\rule{0.850pt}{0.400pt}}
\put(539,624.17){\rule{2.700pt}{0.400pt}}
\multiput(539.00,625.17)(7.396,-2.000){2}{\rule{1.350pt}{0.400pt}}
\multiput(552.00,622.95)(2.472,-0.447){3}{\rule{1.700pt}{0.108pt}}
\multiput(552.00,623.17)(8.472,-3.000){2}{\rule{0.850pt}{0.400pt}}
\put(564,619.17){\rule{2.500pt}{0.400pt}}
\multiput(564.00,620.17)(6.811,-2.000){2}{\rule{1.250pt}{0.400pt}}
\put(576,617.17){\rule{2.500pt}{0.400pt}}
\multiput(576.00,618.17)(6.811,-2.000){2}{\rule{1.250pt}{0.400pt}}
\multiput(588.00,615.95)(2.695,-0.447){3}{\rule{1.833pt}{0.108pt}}
\multiput(588.00,616.17)(9.195,-3.000){2}{\rule{0.917pt}{0.400pt}}
\put(601,612.17){\rule{2.500pt}{0.400pt}}
\multiput(601.00,613.17)(6.811,-2.000){2}{\rule{1.250pt}{0.400pt}}
\multiput(613.00,610.95)(2.472,-0.447){3}{\rule{1.700pt}{0.108pt}}
\multiput(613.00,611.17)(8.472,-3.000){2}{\rule{0.850pt}{0.400pt}}
\put(625,607.17){\rule{2.700pt}{0.400pt}}
\multiput(625.00,608.17)(7.396,-2.000){2}{\rule{1.350pt}{0.400pt}}
\multiput(638.00,605.95)(2.472,-0.447){3}{\rule{1.700pt}{0.108pt}}
\multiput(638.00,606.17)(8.472,-3.000){2}{\rule{0.850pt}{0.400pt}}
\put(650,602.17){\rule{2.500pt}{0.400pt}}
\multiput(650.00,603.17)(6.811,-2.000){2}{\rule{1.250pt}{0.400pt}}
\put(662,600.17){\rule{2.500pt}{0.400pt}}
\multiput(662.00,601.17)(6.811,-2.000){2}{\rule{1.250pt}{0.400pt}}
\multiput(674.00,598.95)(2.695,-0.447){3}{\rule{1.833pt}{0.108pt}}
\multiput(674.00,599.17)(9.195,-3.000){2}{\rule{0.917pt}{0.400pt}}
\put(687,595.17){\rule{2.500pt}{0.400pt}}
\multiput(687.00,596.17)(6.811,-2.000){2}{\rule{1.250pt}{0.400pt}}
\multiput(699.00,593.95)(2.472,-0.447){3}{\rule{1.700pt}{0.108pt}}
\multiput(699.00,594.17)(8.472,-3.000){2}{\rule{0.850pt}{0.400pt}}
\put(711,590.17){\rule{2.700pt}{0.400pt}}
\multiput(711.00,591.17)(7.396,-2.000){2}{\rule{1.350pt}{0.400pt}}
\multiput(724.00,588.95)(2.472,-0.447){3}{\rule{1.700pt}{0.108pt}}
\multiput(724.00,589.17)(8.472,-3.000){2}{\rule{0.850pt}{0.400pt}}
\put(736,585.17){\rule{2.500pt}{0.400pt}}
\multiput(736.00,586.17)(6.811,-2.000){2}{\rule{1.250pt}{0.400pt}}
\put(748,583.17){\rule{2.500pt}{0.400pt}}
\multiput(748.00,584.17)(6.811,-2.000){2}{\rule{1.250pt}{0.400pt}}
\multiput(760.00,581.95)(2.695,-0.447){3}{\rule{1.833pt}{0.108pt}}
\multiput(760.00,582.17)(9.195,-3.000){2}{\rule{0.917pt}{0.400pt}}
\put(773,578.17){\rule{2.500pt}{0.400pt}}
\multiput(773.00,579.17)(6.811,-2.000){2}{\rule{1.250pt}{0.400pt}}
\multiput(785.00,576.95)(2.472,-0.447){3}{\rule{1.700pt}{0.108pt}}
\multiput(785.00,577.17)(8.472,-3.000){2}{\rule{0.850pt}{0.400pt}}
\put(797,573.17){\rule{2.700pt}{0.400pt}}
\multiput(797.00,574.17)(7.396,-2.000){2}{\rule{1.350pt}{0.400pt}}
\multiput(810.00,571.95)(2.472,-0.447){3}{\rule{1.700pt}{0.108pt}}
\multiput(810.00,572.17)(8.472,-3.000){2}{\rule{0.850pt}{0.400pt}}
\put(822,568.17){\rule{2.500pt}{0.400pt}}
\multiput(822.00,569.17)(6.811,-2.000){2}{\rule{1.250pt}{0.400pt}}
\put(834,566.17){\rule{2.500pt}{0.400pt}}
\multiput(834.00,567.17)(6.811,-2.000){2}{\rule{1.250pt}{0.400pt}}
\multiput(846.00,564.95)(2.695,-0.447){3}{\rule{1.833pt}{0.108pt}}
\multiput(846.00,565.17)(9.195,-3.000){2}{\rule{0.917pt}{0.400pt}}
\put(859,561.17){\rule{2.500pt}{0.400pt}}
\multiput(859.00,562.17)(6.811,-2.000){2}{\rule{1.250pt}{0.400pt}}
\multiput(871.00,559.95)(2.472,-0.447){3}{\rule{1.700pt}{0.108pt}}
\multiput(871.00,560.17)(8.472,-3.000){2}{\rule{0.850pt}{0.400pt}}
\put(883,556.17){\rule{2.700pt}{0.400pt}}
\multiput(883.00,557.17)(7.396,-2.000){2}{\rule{1.350pt}{0.400pt}}
\multiput(896.00,554.95)(2.472,-0.447){3}{\rule{1.700pt}{0.108pt}}
\multiput(896.00,555.17)(8.472,-3.000){2}{\rule{0.850pt}{0.400pt}}
\put(908,551.17){\rule{2.500pt}{0.400pt}}
\multiput(908.00,552.17)(6.811,-2.000){2}{\rule{1.250pt}{0.400pt}}
\put(920,549.17){\rule{2.500pt}{0.400pt}}
\multiput(920.00,550.17)(6.811,-2.000){2}{\rule{1.250pt}{0.400pt}}
\multiput(932.00,547.95)(2.695,-0.447){3}{\rule{1.833pt}{0.108pt}}
\multiput(932.00,548.17)(9.195,-3.000){2}{\rule{0.917pt}{0.400pt}}
\put(945,544.17){\rule{2.500pt}{0.400pt}}
\multiput(945.00,545.17)(6.811,-2.000){2}{\rule{1.250pt}{0.400pt}}
\multiput(957.00,542.95)(2.472,-0.447){3}{\rule{1.700pt}{0.108pt}}
\multiput(957.00,543.17)(8.472,-3.000){2}{\rule{0.850pt}{0.400pt}}
\put(969,539.17){\rule{2.700pt}{0.400pt}}
\multiput(969.00,540.17)(7.396,-2.000){2}{\rule{1.350pt}{0.400pt}}
\multiput(982.00,537.95)(2.472,-0.447){3}{\rule{1.700pt}{0.108pt}}
\multiput(982.00,538.17)(8.472,-3.000){2}{\rule{0.850pt}{0.400pt}}
\put(994,534.17){\rule{2.500pt}{0.400pt}}
\multiput(994.00,535.17)(6.811,-2.000){2}{\rule{1.250pt}{0.400pt}}
\put(1006,532.17){\rule{2.500pt}{0.400pt}}
\multiput(1006.00,533.17)(6.811,-2.000){2}{\rule{1.250pt}{0.400pt}}
\multiput(1018.00,530.95)(2.695,-0.447){3}{\rule{1.833pt}{0.108pt}}
\multiput(1018.00,531.17)(9.195,-3.000){2}{\rule{0.917pt}{0.400pt}}
\put(1031,527.17){\rule{2.500pt}{0.400pt}}
\multiput(1031.00,528.17)(6.811,-2.000){2}{\rule{1.250pt}{0.400pt}}
\multiput(1043.00,525.95)(2.472,-0.447){3}{\rule{1.700pt}{0.108pt}}
\multiput(1043.00,526.17)(8.472,-3.000){2}{\rule{0.850pt}{0.400pt}}
\put(1055,522.17){\rule{2.700pt}{0.400pt}}
\multiput(1055.00,523.17)(7.396,-2.000){2}{\rule{1.350pt}{0.400pt}}
\multiput(1068.00,520.95)(2.472,-0.447){3}{\rule{1.700pt}{0.108pt}}
\multiput(1068.00,521.17)(8.472,-3.000){2}{\rule{0.850pt}{0.400pt}}
\put(1080,517.17){\rule{2.500pt}{0.400pt}}
\multiput(1080.00,518.17)(6.811,-2.000){2}{\rule{1.250pt}{0.400pt}}
\put(1092,515.17){\rule{2.500pt}{0.400pt}}
\multiput(1092.00,516.17)(6.811,-2.000){2}{\rule{1.250pt}{0.400pt}}
\multiput(1104.00,513.95)(2.695,-0.447){3}{\rule{1.833pt}{0.108pt}}
\multiput(1104.00,514.17)(9.195,-3.000){2}{\rule{0.917pt}{0.400pt}}
\put(1117,510.17){\rule{2.500pt}{0.400pt}}
\multiput(1117.00,511.17)(6.811,-2.000){2}{\rule{1.250pt}{0.400pt}}
\multiput(1129.00,508.95)(2.472,-0.447){3}{\rule{1.700pt}{0.108pt}}
\multiput(1129.00,509.17)(8.472,-3.000){2}{\rule{0.850pt}{0.400pt}}
\put(1141,505.17){\rule{2.500pt}{0.400pt}}
\multiput(1141.00,506.17)(6.811,-2.000){2}{\rule{1.250pt}{0.400pt}}
\multiput(1153.00,503.95)(2.695,-0.447){3}{\rule{1.833pt}{0.108pt}}
\multiput(1153.00,504.17)(9.195,-3.000){2}{\rule{0.917pt}{0.400pt}}
\put(1166,500.17){\rule{2.500pt}{0.400pt}}
\multiput(1166.00,501.17)(6.811,-2.000){2}{\rule{1.250pt}{0.400pt}}
\put(1178,498.17){\rule{2.500pt}{0.400pt}}
\multiput(1178.00,499.17)(6.811,-2.000){2}{\rule{1.250pt}{0.400pt}}
\multiput(1190.00,496.95)(2.695,-0.447){3}{\rule{1.833pt}{0.108pt}}
\multiput(1190.00,497.17)(9.195,-3.000){2}{\rule{0.917pt}{0.400pt}}
\put(1203,493.17){\rule{2.500pt}{0.400pt}}
\multiput(1203.00,494.17)(6.811,-2.000){2}{\rule{1.250pt}{0.400pt}}
\multiput(1215.00,491.95)(2.472,-0.447){3}{\rule{1.700pt}{0.108pt}}
\multiput(1215.00,492.17)(8.472,-3.000){2}{\rule{0.850pt}{0.400pt}}
\put(1227,488.17){\rule{2.500pt}{0.400pt}}
\multiput(1227.00,489.17)(6.811,-2.000){2}{\rule{1.250pt}{0.400pt}}
\multiput(1239.00,486.95)(2.695,-0.447){3}{\rule{1.833pt}{0.108pt}}
\multiput(1239.00,487.17)(9.195,-3.000){2}{\rule{0.917pt}{0.400pt}}
\put(1252,483.17){\rule{2.500pt}{0.400pt}}
\multiput(1252.00,484.17)(6.811,-2.000){2}{\rule{1.250pt}{0.400pt}}
\put(1264,481.17){\rule{2.500pt}{0.400pt}}
\multiput(1264.00,482.17)(6.811,-2.000){2}{\rule{1.250pt}{0.400pt}}
\multiput(1276.00,479.95)(2.695,-0.447){3}{\rule{1.833pt}{0.108pt}}
\multiput(1276.00,480.17)(9.195,-3.000){2}{\rule{0.917pt}{0.400pt}}
\put(1289,476.17){\rule{2.500pt}{0.400pt}}
\multiput(1289.00,477.17)(6.811,-2.000){2}{\rule{1.250pt}{0.400pt}}
\multiput(1301.00,474.95)(2.472,-0.447){3}{\rule{1.700pt}{0.108pt}}
\multiput(1301.00,475.17)(8.472,-3.000){2}{\rule{0.850pt}{0.400pt}}
\put(1313,471.17){\rule{2.500pt}{0.400pt}}
\multiput(1313.00,472.17)(6.811,-2.000){2}{\rule{1.250pt}{0.400pt}}
\multiput(1325.00,469.95)(2.695,-0.447){3}{\rule{1.833pt}{0.108pt}}
\multiput(1325.00,470.17)(9.195,-3.000){2}{\rule{0.917pt}{0.400pt}}
\put(1338,466.17){\rule{2.500pt}{0.400pt}}
\multiput(1338.00,467.17)(6.811,-2.000){2}{\rule{1.250pt}{0.400pt}}
\put(1350,464.17){\rule{2.500pt}{0.400pt}}
\multiput(1350.00,465.17)(6.811,-2.000){2}{\rule{1.250pt}{0.400pt}}
\multiput(1362.00,462.95)(2.695,-0.447){3}{\rule{1.833pt}{0.108pt}}
\multiput(1362.00,463.17)(9.195,-3.000){2}{\rule{0.917pt}{0.400pt}}
\put(1375,459.17){\rule{2.500pt}{0.400pt}}
\multiput(1375.00,460.17)(6.811,-2.000){2}{\rule{1.250pt}{0.400pt}}
\multiput(1387.00,457.95)(2.472,-0.447){3}{\rule{1.700pt}{0.108pt}}
\multiput(1387.00,458.17)(8.472,-3.000){2}{\rule{0.850pt}{0.400pt}}
\put(1399,454.17){\rule{2.500pt}{0.400pt}}
\multiput(1399.00,455.17)(6.811,-2.000){2}{\rule{1.250pt}{0.400pt}}
\multiput(1411.00,452.95)(2.695,-0.447){3}{\rule{1.833pt}{0.108pt}}
\multiput(1411.00,453.17)(9.195,-3.000){2}{\rule{0.917pt}{0.400pt}}
\put(1424,449.17){\rule{2.500pt}{0.400pt}}
\multiput(1424.00,450.17)(6.811,-2.000){2}{\rule{1.250pt}{0.400pt}}
\end{picture}
\end{figure}
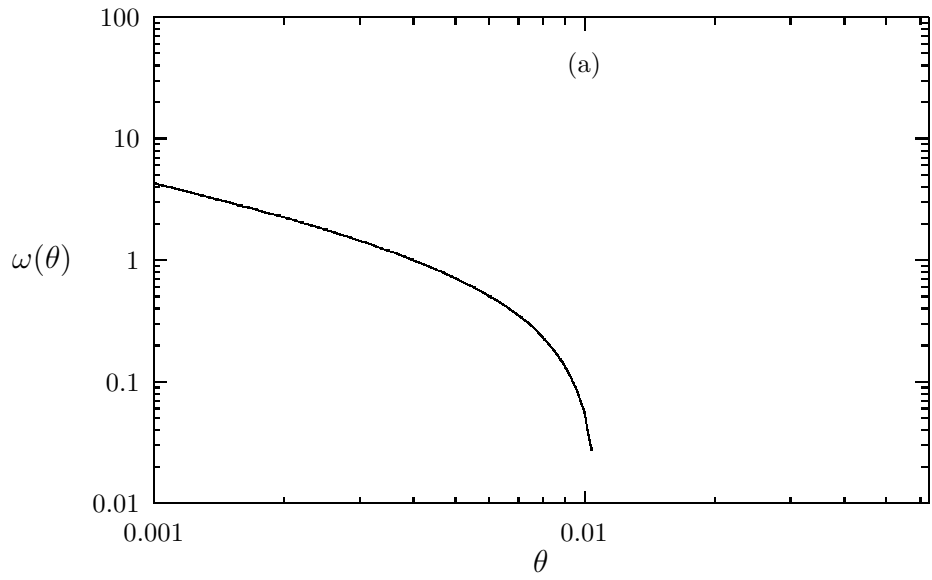
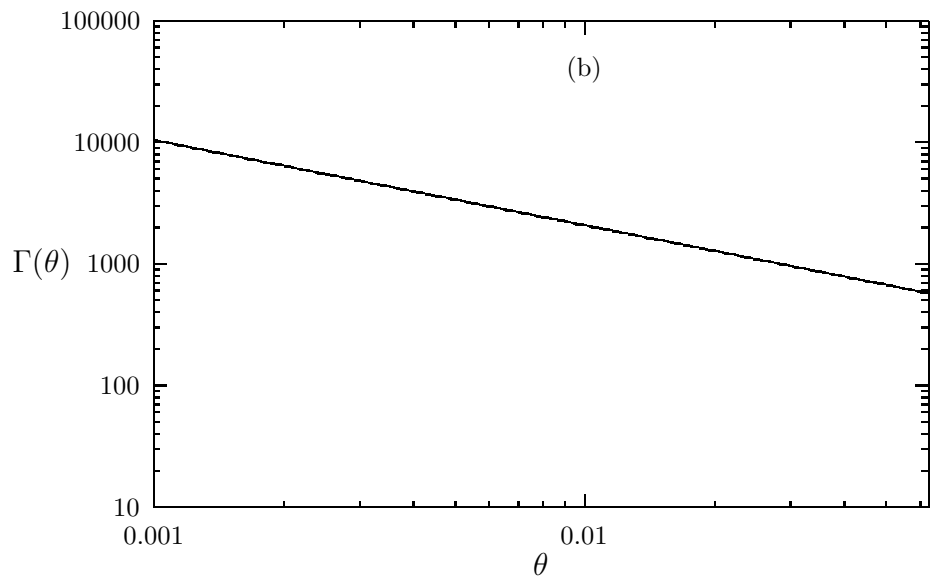
\end{document}